\documentclass[prb,twocolumn,showpacs,superscriptaddress]{revtex4-1}

\usepackage{graphicx}
\begin{document}

\def \reofour {$\mathrm{(TMTSF)_{2}ReO_4}$\,}
\def \clofour {$\mathrm{(TMTSF)_{2}ClO_4}$\,}
\def \areofour {$\mathrm{ReO_4^-}$\,}
\def \aclofour {$\mathrm{ClO_4^-}$\,}
\def \pfsix {$\mathrm{(TMTSF)_{2}PF_6}$\,}
\def \asfsix {$\mathrm{(TMTSF)_{2}AsF_{6}}$}
\def \tmtsfx {$\mathrm{(TMTSF)_{2}X}$\,}
\def \tm2x {$\mathrm{(TM)_{2}X}$\,}

\def\tc{$T_{c}$\,}
\def\pc{$P_{c}$\,}

\def\t6as{$\mathrm{(TMTSF)_{2}AsF_{6}}$}
\def\tmx{$\mathrm{(TMTSF)_{2}(ClO_{4})_{(1-x)}ReO_{4}_x}$\,}
\def\tmc{$\mathrm{(TMTSF)_{2}ClO_{4}}$\,}
\def\tms{$\mathrm{(TMTSF)_{2}AsF_{6(1-x)}SbF_{6x}}$}\,
\def\tmps{$\mathrm{(TMTTF)_{2}PF_{6}}$\,}
\def\tmttfsbf6{$\mathrm{(TMTTF)_{2}SbF_{6}}$\,}
\def\tmttfasf6{$\mathrm{(TMTTF)_{2}AsF_{6}}$\,}
\def\tmttfbf4{$\mathrm{(TMTTF)_{2}BF_{4}}$\,}
\def\tmtsfreo4{$\mathrm{(TMTSF)_{2}ReO_{4}}$\,}
\def\tq{$\mathrm{TTF-TCNQ}$\,}
\def\tsq{$\mathrm{TSeF-TCNQ}$}\,
\def\qnq{$(Qn)TCNQ_{2}$}\,
\def\R{$\mathrm{ReO_{4}^{-}}$}  
\def\C{$\mathrm{ClO_{4}^{-}}$}
\def\P{$\mathrm{PF_{6}^{-}}$}
\def\tqr{$\mathrm{TCNQ^\frac{\cdot}{}}$\,}
\def\nmpq{$\mathrm{NMP^{+}(TCNQ)^\frac{\cdot}{}}$\,}
\def\q{$\mathrm{TCNQ}$\,}
\def\nmp{$\mathrm{NMP^{+}}$\,}
\def\f{$\mathrm{TTF}\,$}
\def\tc{$T_{c}$\,}
\def\nmq{$\mathrm{(NMP-TCNQ)}$\,}
\def\ts{$\mathrm{TSeF}$}
\def\tsm{$\mathrm{TMTSF}$\,}
\def\tst{$\mathrm{TMTTF}$\,}
\def\tmp6{$\mathrm{(TMTSF)_{2}PF_{6}}$\,}
\def\tms2x{$\mathrm{(TMTSF)_{2}X}$}
\def\as{$\mathrm{AsF_{6}}$}
\def\sb{$\mathrm{SbF_{6}}$}
\def\pf{$\mathrm{PF_{6}}$}
\def\re{$\mathrm{ReO_{4}}$}
\def\ta{$\mathrm{TaF_{6}}$}
\def\cl{$\mathrm{ClO_{4}}$}
\def\4fb{$\mathrm{BF_{4}}$}
\def\ttdm{$\mathrm{(TTDM-TTF)_{2}Au(mnt)_{2}}$}
\def\edt{$\mathrm{(EDT-TTF-CONMe_{2})_{2}AsF_{6}}$}
\def\tfx{$\mathrm{(TMTTF)_{2}X}$\,}
\def\tsx{$\mathrm{(TMTSF)_{2}X}$\,}
\def\ttftcnq{$\mathrm{TTF-TCNQ}$\,}
\def\ttf{$\mathrm{TTF}$\,}
\def\tcnq{$\mathrm{TCNQ}$\,}
\def\bedtttf{$\mathrm{BEDT-TTF}$\,}
\def\reo4{$\mathrm{ReO_{4}}$}
\def\bedtttfreo4{$\mathrm{(BEDT-TTF)_{2}ReO_{4}}$\,}
\def\et2i3{$\mathrm{(ET)_{2}I_{3}}$\,}
\def\et2x{$\mathrm{(ET)_{2}X}$\,}
\def\ket2x{$\mathrm{\kappa-(ET)_{2}X}$\,}
\def\cuncnbr{$\mathrm{Cu(N(CN)_{2})Br}$\,}
\def\ket2x{$\mathrm{\kappa-(ET)_{2}X}$\,}
\def\cuncncl{$\mathrm{Cu(N(CN)_{2})Cl}$\,}
\def\cuncs{$\mathrm{Cu(NCS)_{2}}$\,}
\def\betsfecl4{$\mathrm{(BETS)_{2}FeCl_{4}}$\,}
\def\bets{$\mathrm{BETS}$\,}
\def\hc2{$H_{c2}(T)$}
\def\et{$\mathrm{ET}$\,}
\def\tmm{$\mathrm{TM}$\,}
\def\tmtsf{$\mathrm{(TMTSF)}$\,}
\def\tmttf{$\mathrm{(TMTTF)}$\,}
\def\tm2x{$\mathrm{(TM)_{2}X}$\,}
\def\t1{${1/T_1}$\,}
\title{Domain walls at the spin density wave endpoint of the organic superconductor \pfsix under pressure}
\author {N. Kang}
\author {P. Auban-Senzier}
\author {D. J\'erome}
\author {C.R. Pasquier}
\affiliation{Laboratoire de Physique des Solides, UMR 8502-CNRS, Univ.Paris-Sud, Orsay, F-91405, France}
\author {B. Salameh}
\affiliation{Laboratoire de Physique des Solides, UMR 8502-CNRS, Univ.Paris-Sud, Orsay, F-91405, France}
\affiliation{Department of Applied Physics, Tafila Technical University, Tafila, Jordan}
\author{S. Brazovskii}
\affiliation{LPTMS-CNRS, UMR 8626, Univ.Paris-Sud Bat 100, Orsay, F-91405, France }


\begin{abstract}

We report a comprehensive investigation of the organic superconductor \pfsix in the vicinity of the endpoint of the spin density wave - metal phase transition where phase coexistence occurs. At low temperature, the transition of metallic domains towards superconductivity is used to reveal the various textures. In particular, we demonstrate experimentally the existence of 1D and 2D metallic domains with a cross-over from a filamentary superconductivity mostly along the $c^{\star}$-axis to a 2D superconductivity in the $b'c$-plane perpendicular to the most conducting direction. The formation of these domain walls may be related to the proposal of a soliton phase in the vicinity of the critical pressure of the \tmp6 phase diagram.
\end{abstract}

\pacs{73.61.-r, 73.23.-b, 73.50.-h}
\maketitle
Understanding the evolution from a  magnetically ordered metallic (possibly insulating) ground state to a paramagnetic and metallic (M) (potentially superconducting) ground state is a long standing problem in condensed matter physics. Such a situation is encountered  in very diverse systems such as heavy fermion compounds, cuprates, and the recently discovered pnictide superconductors. In all these systems the parameter controlling the phase stability can be a dopant concentration, pressure or magnetic field. 
Pressure was also at the origin of the discovery of superconductivity (SC) in the  quasi one dimensional charge transfer salt, \tmp6, where  an insulating itinerant antiferromagnetic phase known as a spin density wave (SDW) ground state is stabilized at low temperature through a second order phase transition. As the magnetic order can be driven to zero temperature by pressure with the stabilization of SC above \pc $\approx$ 9 kbar, one would  be entitled to believe that the  \tmp6 phase diagram  provides a good experimental playground for the study of a SDW quantum critical point. The study of the border region between SDW and SC becomes therefore an important issue for organic superconductivity since no consensus exists yet regarding the pairing mechanism and there has been a proposal for a microscopic coexistence of magnetic and superconducting order in a narrow pressure domain implying non nested region on the Fermi surface in the vicinity of the boundary \pc \cite{Yamaji82}. Early studies\cite{Greene80,Brusetti82} have recognized  that the transition from the SDW to the metallic state is of first order in this pressure regime which has been in turn extensively revisited by various techniques in the last decade. Resistivity measurements were performed by Vuletic \textit{et al.}\cite{Vuletic02} making small pressure increments up to \pc and subsequently by Kornilov\textit{et al.} \cite{Kornilov04} at a fixed pressure but monitoring the distance to \pc \textit{via} an applied magnetic field. Both studies concluded to the coexistence of the two phases SDW/M or SDW/SC although in spatially separated regions. The possibility of metallic slabs becoming superconducting at low temperature in the pressure regime where \tc remains constant was suggested by transport data along the most conducting axis and also supported by a drastic enhancement of the upper critical field \cite{Lee02,Colin08,Ruetschi09}. Furthermore, Vuletic \textit{et al.}\cite{Vuletic02} pointed out the existence of a particular pressure, $P_{c0}$, related to a sudden vanishing of SC coherence.  Simultaneous measurements of NMR and transport at a given pressure have corroborated the claims for macroscopic coexistence coming from transport data and have also provided an analysis of the volume fraction as a function of temperature\cite{Yu02, Lee05}. However, the comprehensive pressure mapping of this coexistence regime SDW-M(SC) in the $P-T$ phase diagram is still missing as well as how the minority phase M self-organizes within the majority SDW phase. On theoretical grounds, various approaches have been developed: Ginzburg-Landau like models have succeeded to obtain a phase coexistence between SC and SDW states \cite{Podolsky04} and a modulation of the SC and SDW order parameters along both $a$ and $b$ axes has been suggested \cite{Zhang06}. A microscopic approach has also been developed \cite{soliton,Gorkov05} based on the soliton theory which leads to a modulation of the SC and SDW order parameters along the $a$-axis.
\begin{figure}[htb]
	\centering
		\includegraphics[width=0.85\hsize]{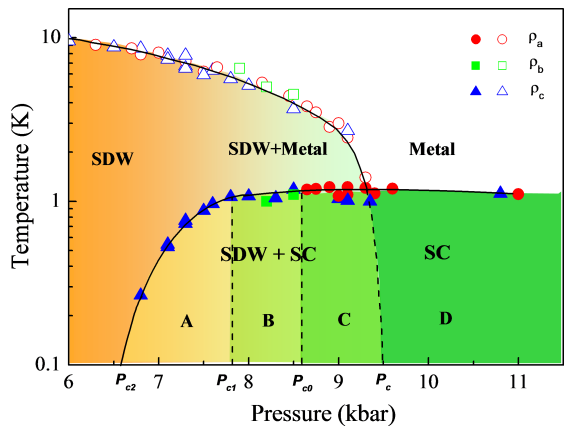}
		\caption{(Color online) Phase diagram of \pfsix as determined from resistivity measurements along the three axes (circles: $\rho_{a}$; squares: $\rho_{b}$; triangles: $\rho_{c}$ ). The filled (open) symbols correspond to the transition towards SC (SDW) respectively. The contrast of colors between $P_{c2}$ and $P_{c}$ illustrates the increase in SC volume fraction from $P_{c2}$ up to  $P_{c}$ corresponding to the three different regimes explained in the text. Based on the knowledge of $T_{SDW}$ and \tc, the pressure of 5.5 kbar in ref\cite{Lee05} would correspond to 8.8 kbar with the present pressure scale.}
	\label{Phase}
\end{figure}

In this paper, we explore the emergence of the minority phase, metallic (or SC at low temperature) from the pure SDW state and how it evolves towards the homogeneous metal (or SC) state under pressure.  We use superconductivity as a tool to decorate the texture by comparing the temperature dependence of resistivity experiments performed along the $a$, $b'$ and $c^\star$ axes. This texture is in favor of the soliton model.

Resistivity measurements were performed in high-quality \tmp6 single crystals from the batch used in an earlier study \cite{Vuletic02}. Gold plated electrical contacts were evaporated on the sample surfaces to measure $\rho_{a}$, $\rho_{b}$ and $\rho_{c}$ along $a$, $b'$ and $c^\star$ axes respectively on different samples. The resistance measurements were performed using a standard low frequency lock-in detection. The applied current was chosen in order to remain below the SC critical current along the considered axis for each pressure and to minimize heating effects. The measurements were carried out in a dilution refrigerator ($T\geq 50mK$) with a magnetic field always applied along the $c^\star$-axis. Measuring the resistivity tensor on the same sample would have obviously been the most satisfactory solution but this happens to be non feasible. Indeed, contacts evaporated on the crystal surfaces for the measurement along a given axis always short circuit and consequently preclude measurements along a perpendicular axis. 
For a comparison of the transport anisotropy at a given pressure $P$ we chose among our various pressure runs the ones corresponding to $P\pm 0.1$ kbar. Hydrostatic pressures up to $11$ kbar were generated by using a Be-Cu clamp cell with Daphn\'e silicon oil as the pressure transmitting medium. Given the importance to study transport along the three axes at the same pressure, a determination of the pressure or at least of the relative pressure between different runs is of crucial importance for the present study. This was achieved at low temperature using as an \textit{in situ} pressure gauge, the pressure dependence of the sharp SDW transition reported in Vuletic \textit{et al.}\cite{Vuletic02}. 
The main result of this new study is the establishment of a detailed phase diagram for the coexistence region which is displayed on Fig.\ref{Phase}. As shown in this Fig.\ref{Phase}, the domain of the \tmp6 phase diagram where SC is observed can be subdivided into four different regions according to the response of transport to SC along the different axes. In particular, the SDW/M(SC) phase coexistence is observed between $P_{c2}=6.6$ kbar and $P_c=9.4 $kbar with a strong increase of the critical temperature between $P_{c2}$ and $P_{c1}=7.8$kbar (phase A), a much weaker one between $P_{c1}$ and $P_{c0}$ (phase B) and finally, \tc remains pressure independent above $P_{c0}$ (phase C).

\begin{figure}[htb]
	\centering
	\includegraphics[width=0.85\hsize]{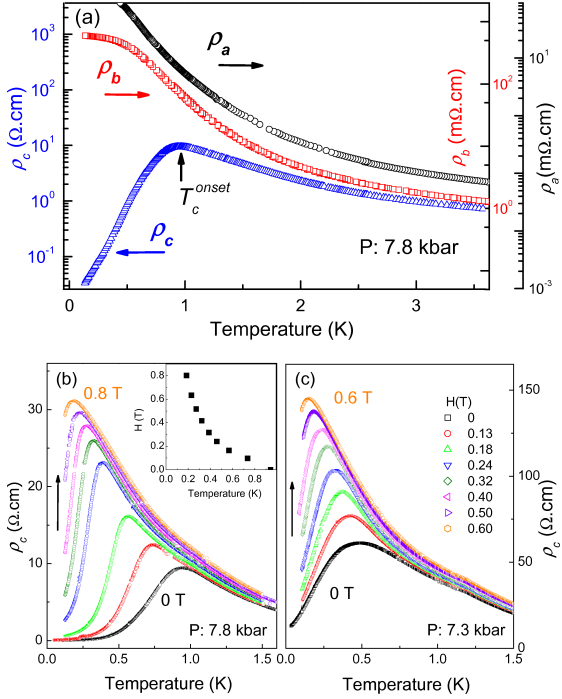}
	\caption{(Color online) \textbf{Phase A}: (a) Temperature dependence of $\rho_{a}$, $\rho_{b}$ and $\rho_{c}$ at $P=7.8$ kbar. (b) Temperature dependence of $\rho_{c}$ at $P=7.8$ kbar for magnetic fields ranging from $0$ to $0.8T$ by step of $0.1T$. The insert shows the deduced upper critical field line. (c) Temperature dependence of $\rho_{c}$ at $P=7.3$ kbar for different magnetic fields.}
	\label{Figure2}
\end{figure}

\textbf{Phase A}, $P_{c2}=6.6 <P< P_{c1}=7.8$ kbar: as shown in Fig.\ref{Figure2}a, while the resistivities along the three axes exhibit similar insulating temperature dependences for $T>1K$, only $\rho_{c}$ exhibits a partial SC transition. In contrast, $\rho_{a}(T)$ exhibits the same insulating behavior as in the low pressure purely SDW state over the whole measured $T$ range. $\rho_{b}(T)$ follows $\rho_{a}(T)$ except near $P_{c1}$ where it exhibits a saturation at low temperatures. The $T_c^{onset} (P)$, in Fig.\ref{Phase}, is defined by the onset of superconductivity namely, the maximum of $\rho_{c}(T)$ at a given pressure, see Fig.\ref{Figure2}a. The sensitivity of SC to magnetic field is shown in Fig. \ref{Figure2}(b) and (c) by the evolution of $\rho_{c}(T)$ with the applied magnetic field at $P=7.3$ and $P=7.8$ kbar. The upward curvature of the upper critical field down to the lowest temperatures is in agreement with previous reports in \tm2x salts \cite{Lee02, Colin08, Ruetschi09}. In this phase A, a higher pressure increases \tc and reduces the broadness of the transition. Such a behavior is typical of phase separation as long as SC domains are smaller than the penetration depth. Our data are also compatible with the formation of filaments elongated mainly along the $c^{\star}$-axis which may cross the whole thickness of the sample approaching $P_{c1}$. Indeed, our observations looks qualitatively similar to the results for SC wires \cite{Bezryadin00} where the inherent presence of phase slips give rise to finite resistance below $T_{c}$. 
 
\begin{figure}[htb]
	\centering
	\includegraphics[width=0.85\hsize]{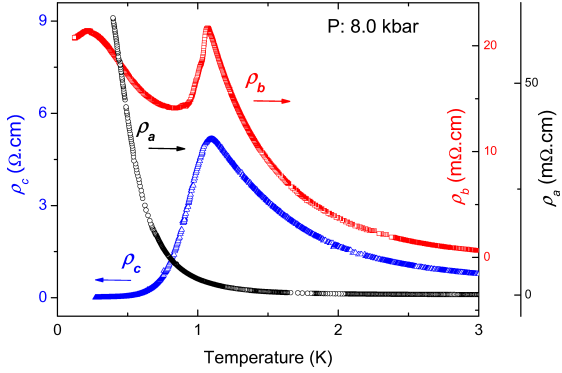}
	\caption{(Color online) \textbf{Phase B, P=8.0kbar}: temperature dependence of $\rho_{a}$, $\rho_{b}$ and $\rho_{c}$ at zero magnetic field.}
\label{Figure3}
\end{figure}
\begin{figure}[htb]
\centerline{\includegraphics[width=0.85\hsize]{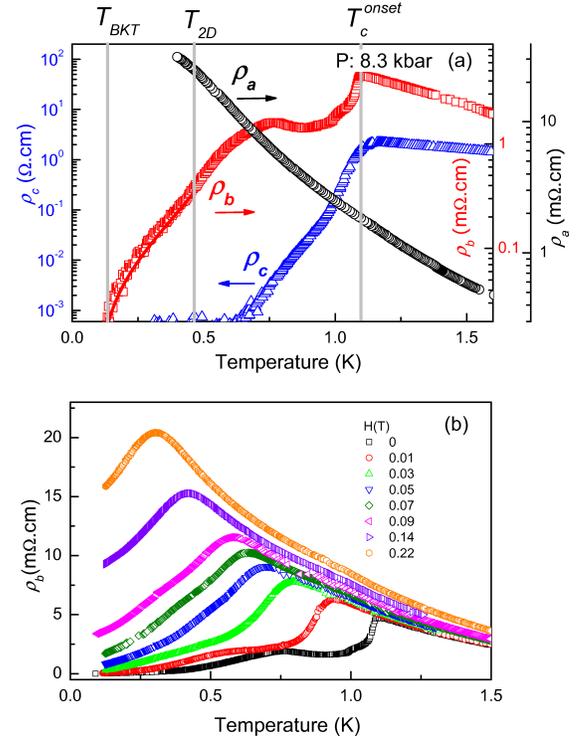}}
\caption{(Color online) \textbf{Phase B, P=8.3kbar}: (a) Temperature dependence of $\rho_{a}$, $\rho_{b}$ and $\rho_{c}$. The line through the data points of  $\rho_{b}$ corresponds to the fit of $\rho_{b}$ by the BKT model. (b) Temperature dependence of $\rho_{b}$ for different magnetic fields.}
\label{Figure4}
\end{figure}

\textbf{Phase B}, $P_{c1}=7.8<P<P_{c0}=8.6$ kbar: as shown in Fig.\ref{Figure3} and \ref{Figure4}, both $\rho_{b}(T)$ and $\rho_{c}(T)$ exhibit a SC transition. The drop of $\rho_{b}$ to a finite resistance state reproduces the broad decrease of $\rho_{c}(T)$ at $T_c^{onset} (P)$ and can be attributed to the SC transition in the metallic domains, coexisting with the SDW background. At lower temperatures, the increase of $\rho_{b}(T)$ infers that SDW domains are in series with SC domains along $b'$. A (true) zero resistance state along $c^{\star}$-axis is achieved, in phase B, at a temperature which increases with pressure. However, at both $P=8.0$ kbar and $P=8.3$ kbar, $\rho_{a}$ still remains insulating. Therefore, the system looks like an array of SC-SDW-SC junctions with Josephson coupling across insulating barriers, all located in $b'c$-planes. The in-plane Josephson coupling increases with higher pressure or lower temperature leading to superconducting correlation in the $b'$ direction and in turn to 2D SC within $b'c$-planes. Hence, below $\sim0.2K$, an enhanced Josephson coupling allows a weak decrease of $\rho_{b}(T)$ at $P=8.0$ kbar which shifts to larger temperatures upon increasing pressure, that is $\sim0.76K$ at $P=8.3$ kbar. The existence of SC along the $b'$ axis is confirmed by the disappearance of SC under a finite magnetic field as shown in Fig.\ref{Figure4}(b). This phenomenon is typical of granular superconductors and superconductor-insulator transition systems \cite{Goldman} and disappears at $P_{c0}$ where both $\rho_{b}(T)$ and $\rho_{c}(T)$ present a (single) sharp transition at $T_c^{onset} (P)$. The 2D nature of SC in phase B is confirmed on Fig.\ref{Figure4}(a) by the fit of the $\rho_{b}(T)$ curve below $T_{2D}\sim0.45 K$, by a model considering a 2D SC above its Berezinskii-Kosterlitz-Thouless (BKT) transition temperature, $T_{BKT}\sim0.15 K$, where the resistance reads,
$R_{BKT}(T)=R_{0}\,exp -\left(G_i^{2D}\frac{T_{BKT}}{T-T_{BKT}}\right)^{1/2}\label{BKTR},
$
where $G_i^{2D}\sim T_{BKT}/\sqrt{t_b t_c}$ is the 2D Ginzburg parameter and $R_{0}$, a fitting parameter \cite{Tinkham}.

\textbf{Phase C}, $P_{c0}=8.6 <P<P_{c}= 9.4$ kbar: both $\rho_{b}(T)$ and $\rho_{c}(T)$ present a sharp transition at $T_c^{onset} (P)$ with a zero resistance state below $T_c^{onset} (P)$. $\rho_{a}(T)$ data have been already presented \cite{Vuletic02}: the pressure evolution of $\rho_{a}$ mimics the evolution of $\rho_{b}$ in phase B. In particular, a 'double transition' in $\rho_{a}(T)$ is observed nearly up to $P_c$.

\textbf{Phase D}: superconductivity appears to be homogeneous above \pc = 9.4 kbar.
The starting frame of any interpretation is the electronic zone in the reciprocal lattice with the electronic spectrum $E(\vec{k})$ satisfying the nesting condition $E(\vec{k}+\vec{Q})\approx -E(\vec{k})$ (with the accuracy of $\Delta$ since at low $T$ the state is insulating).

The commonly used model limits the major spectrum also to only nearest neighbors overlaps: $E(\vec{k})=-2t_{a}\cos k_{a}-2t_{b0}\cos k_{b}$ leading to the common-sense nesting wave number $\vec{Q_{0}}=2\pi(1/2,1/2,1/2)$. (The wave numbers, $k_{i}$, are taken in units of inverse lattice parameters.) But the SDW was always recognized to be incommensurate, and moreover its wave number has been well determined, in $a$ and $b$ directions, as $\vec Q_{SDW}=2\pi(1/2,q_{b},q_{c})$ - with $q_{b}=1/4\pm0.05$, not $1/2$! These direct X-ray results \cite{X-ray} agree with simulations from the NMR studies \cite{Delrieu:86,Takahashi:86} giving $q_{b}$ as $0.2$ or $0.3$. That was elucidated by band structure calculations  \cite{Ducasse:85} as an ill-expected interference of oblique inter-stack overlaps, $t_{b1}$ between the nearest molecular stacks in $b$ direction, but among molecules which are next nearest neighbors along the stack: 
$E(\vec{k})=-2t_{a}\cos k_{a}-2t_{b0}\cos k_{b}\ -2t_{b1}\cos(k_{b}-k_{a})$. 
Having written it \cite{Yamaji82}, at the Fermi sheets $k_{a}\approx\pm\pi/2$, as $E(\vec{k})=\pm v_{F}\delta k_{a}-2t_{b}(k_{a})\cos(k_{b}\mp\Phi_{0})$, $\Phi_{0}=\pm\arctan(t_{b1}/t_{b0})$, one sees that the interference does not destroy the nesting but shifts its vector, in $b$ direction, from $\pi$ to  $q_{b}=\pi-2\Phi_{0}$. For room temperature crystal parameters the effect is small as expected, but, at low $T$, it becomes as large \cite{Ducasse:85} as to shift $q_{b}$ from $1/2$ to the vicinity of $1/4$.

The metalization and progressive destruction of the SDW state is determined by the antinesting energy $E_{anti}(\vec{k})=(E^{\prime}(\vec{k})+E^{\prime}(\vec{k}+\vec{Q}))/2$. It is given by the smaller contributions $E^{\prime}(\vec{k})=-2t_{c}\cos k_{c}-2t_{b}^{\prime}\cos2k_{b}$ considering them at the new nesting vector $\vec{Q}$ as it is determined by the dominant term. The conventional candidate for unnesting, $-2t_{b}^{\prime}\cos2k_{b}$, gives $E_{anti}^{b}(\vec{k})=-t_{b}^{\prime}(\cos 2k_{b}+\cos(2k_{b}+4\pi q_{b}))$. For the commonly supposed $q_{b}=1/2$, the two terms are identical giving $E_{anti}^{b}(\vec{k})=-2t_{b}^{\prime}\cos2k_{b}$. But now, for $q_{b}=1/4$, the two terms have opposite signs, so $E_{anti}^{b}(\vec{k})$ just vanishes. Although $q_b$ may not be exactly $1/4$, the incommensurability of the SDW induces a noticeable decrease of $E_{anti}^{b}(\vec{k})$. Also, the effect of oblique overlaps slightly decrease with pressure \cite{Ducasse:85}, hence the compensation of unnesting in $b$ direction reduces and this direction starts to play a bigger role. That seems to correlate with our observations. The $c$-axis term, $E_{anti}^{c}(\vec{k})=-t_{c}(\cos k_{c}+\cos(k_{c}+2\pi q_{c}))$, survives: even if $q_{c}$ is not well determined, for all data $q_{c}\neq1/2$\cite{X-ray,Takahashi:86} -  there are no major terms to fix it as it was for $q_b$. Therefore, most functions of the SDW destruction, formation of the solitonic midgap state or of spill-over pockets, and finally of stabilization of initially fragmented solitonic walls, - all are maintained by electronic hybridization in the nominally weakest $c$ direction. This picture is coherent with our observation, at first sight counterintuitive, that the SC develops first in the direction of worst conduction.


In conclusion, we have reported the first comprehensive investigation of the coexistence region in the pressure-temperature phase diagram of \tmp6 near the critical pressure \pc, in which the SC phase is inhomogeneous and spatially modulated. This regime is characterized by conducting (SC) slabs perpendicular to the most conducting axis which originate from the coalescence of metallic domains elongated mainly along the $c^{\star}$-axis at low pressure as evidenced from the onset of superconductivity first along $c^{\star}$ while the system remains insulating along the perpendicular directions. At increasing pressure, metallic (SC) coherence sets in along the $b'$ direction as well. An improvement of the model, coherent to both new and old overlooked observations, is proposed to understand the counterintuitive experimental picture. Our study might be extended in the SDW/M regime above \tc as already suggested \cite{Lebed84}, even if the texture is more difficult to extract in this regime. The existence of a textured SC phase at the border of the SDW/metal transition in \pfsix could help to shed new light on the nature of coexistence of two ordered phases in other strongly correlated systems, other \tm2x salts as well as the recently discovered iron-pnictide superconductors\cite{pnictides}.

\begin{acknowledgments}

This work was supported by the European Community under the project CoMePhS (grant no. NMPT4-CT-2005-517039). N. K. and B.S. also acknowledge financial support from this grant. S.B. acknowledges supports of the ANR program (project BLAN07-3-192276).
\end{acknowledgments}


\end{document}